\documentclass[onecolumn,preprintnumbers,amsmath,amssymb,showkeys]{revtex4}
\usepackage{graphicx}% Include figure files
\usepackage{dcolumn}% Align table columns on decimal point
\usepackage[utf8x]{inputenc}
\usepackage{bm}% bold math
\usepackage{float}
\begin{document}

\title{Mass Spectra and Decay Constants of Heavy-Light Mesons: A Case Study of QCD Sum Rules and Quark Model}% Force line breaks with \\

\author{Halil Mutuk}
\email{halilmutuk@gmail.com}
 \affiliation{Physics Department, Ondokuz Mayis University, 55139, Samsun, Turkey}%Lines break automatically or can be forced with \\
 


\begin{abstract}
In this paper we visited mass spectra and decay constants of pseudoscalar and vector heavy-light mesons ($B$, $B_s$, $D$, and $D_S$) in the framework of QCD sum rule and quark model.  The harmonic oscillator wave function was used in quark model while a simple interpolating current was used in QCD sum rule calculation.  We obtained good results in accordance with the available experimental data and theoretical studies.
\end{abstract}

\maketitle

\section{\label{sec:level1}Introduction}

The ultimate objective of particle physics is to investigate and examine the structure and the origin of matter. For this purpose many theoretical and experimental endeavors are made, and a resulting model was theorized, which we call the Standard Model of particle physics. Quark model which was proposed by Gell-Mann and Zweig in 1964 \cite{1} is a part of the Standard Model, and interprets hadrons fairly compatible with the experimental data. According to the quark model, mesons are made of quark-antiquark pairs ($q\bar{q}$) and baryons are made of three quarks $qqq$ or antiquarks $\bar{q}\bar{q}\bar{q}$. These quarks interact with each other via emitting and/or absorbing gluons. The resulting theory which explains these interactions is is the Quantum Chromodynamics (QCD). 

 The interaction of quarks is described by QCD, which is part of the Standart Model of particle physics. QCD is thought to be the {\it true} theory of strong interactions. QCD is a SU(3) gauge theory describing the interactions of six quarks which transform under the fundamental representation of SU(3) group via the exchange of gluons, that transform under the adjoint representation. Although it has been more than 50 years that QCD has been proposed, a solution has been evaded. Contrary to electroweak theory, where it is possible to obtain precise results using perturbation theory, the order of precision obtained in QCD has been lower by orders of magnitude. The main reason for this is that the coupling constant  (which should be the perturbation parameter) of QCD is of the order one in low energies, hence the truncation of the perturbative expansion can not be carried out. However, it is an important subject to study the spectrum of particles predicted by QCD.

Since perturbation theory is not applicable, a non-perturbative approach has to be used to study systems that involve strong interactions. Some of the non-perturbative approaches to strongly interacting systems are the QCD sum rules, quark models and lattice QCD. The advantage of QCD sum rules and lattice QCD is that they are based on QCD itself, whereas in quark models, one assumes a potential energy between the quarks and solve a Schrödinger-like equation. The advantage of quark models, on the other hand, is that it allows one to study also the excited states, whereas in QCD sum rules and lattice QCD, only the ground state or in some exceptional cases the first excited state can be studied. 

 In quark models one assumes a potential interaction among quarks which makes model as a non-relativistic approach. Therefore, the systems that are best suited for study in quark models are the heavy quark system which contain $c$ or $b$ quarks. The bare masses of $u$, $d$ and $s$ quarks are 2 MeV, 4 MeV and 96 MeV, respectively \cite{2}. At a first look, quark model seems rather difficult to apply to light quarks. Capstick et al. presented reasonable explanations to link quark models including a  minimal amount of relativity to the basics of QCD \cite{3}. Although the pole masses of $u$, $d$ and $s$ quarks are very low and hence they are relativistic, in constituent quark models, instead of treating the physical $u$, $d$ and $s$ quarks, one treats the so called constituent quarks, which are nothing else than quarks dressed by gluons and other sea quarks inside the hadron. The masses of constituent quarks are around 300 MeV and hence they can also be treated in non-relativistic quark models. Such an approach has been applied to light quark systems with a surprising success \cite{4,5,6}, leading to that model so-called Constituent Quark Model (CQM), which -based on the Gell Mann-Zweig idea- explains meson and baryon bound systems. 

Different situation is for heavy-light quark systems $(Q\bar{q})$.  For example an electron is more relativistic in the hydrogen atom $(p,e^-)$ than in the positronium atom $(e^+e^-)$ \cite{7}. Positronium can be taken as a naive model for quarkonium. The binding energy of the positronium is half of the hydrogen atom, and is small compared to the electron mass. For this reason the positronium bound state can be described by non-relativistic quantum mechanics. But the decay of the positronium resonance is a purely relativistic phenomenon. Nevertheless, we can attempt to apply the quark model to heavy-light mesons. The outcome of this attempt is the not directly using of Heavy Quark Symmetry  (HQS), but one aspect of it. Mesons are two particle systems and the reduced mass is dominated by the light quark mass, $\frac{1}{\mu}=\frac{1}{m}+\frac{1}{M} \simeq \frac{1}{M}$ if $M>>m$. The spectra for $(c\bar{q})$ and $(b\bar{q})$  should be very similar under this assumption \cite{7}. Indeed reasonable spectroscopy of $D$ and $B$ mesons can be obtained. There is a rich literature for the spectrum and dynamics of the heavy-light mesons, for example \citep{8,9,10,11,12,13,14,15,16,17,18,19,20,21,22,23,24,25}. In \cite{26}, they studied semileptonic $D$ and $D_s$ decays based on the predictions of the relevant form factors from the covariant light-front quark model. In \cite{27}, the authors studied the Cabibbo-Kobayashi-Maskawa matrix element $\vert V_{ub} \vert$ which is not determined up to now in inclusive or exclusive $B$ decays.

Light quark physics is a key topic to understand the nature of QCD. They can be thought of a probe of the strong interactions by means of non-perturbative effects \cite{28}. Heavy-light meson systems $(Q\bar{q})$ is also central to enlighten the nature of QCD and strong interactions. Heavy-light meson spectroscopy has been the subject of both theoretical and experimental studies since the 2000s. Especially in the charm sector, new excited states were observed in $D$ and $D_s$ mesons \citep{28,29,30,31,32}. 

An important feature of B meson physics is that it is sensitive to New Physics (NP) Beyond the Standard Model (BSM) via rare decays. Furthermore hadronic decay channels of B mesons might have more systematic uncertainties due to the model indetermination, compared to the lepton/photon decay channels. Thus studying  $B \to lepton / photon $ decays present a play field for the search of NP. Beside that, B factory experiments BaBar and Belle were built to test the description of quark mixing in the Standard Model.  The first theoretical description of quark mixing was proposed by Cabibo in 1963 \cite{33}. One year later in 1964, Christenson et al.  discovered  CP violation in neutral kaon decays with a tiny friction \cite{34}. This phenomena is referred as to conclusion that matter and antimatter might behave differently. This phenomena is referred as to conclusion that matter and antimatter can behave differently. Kobayashi and Maskawa generalized Cabibbo's idea by adjusting new quarks to the model \cite{35}. In the framework of Standard Model, CP violation can be accommodated by introducing a complex phase in the $3 \times 3 $ unitary Cabibo-Kobayashi-Maskawa (CKM) matrix. Indeed this phase can be measured in experiments. The cost of adding a parameter is to use a third generation of quarks.  CP violation also occurs in B  decays. The B factories were built to test for this purpose. B factories gave a substantial contribution to particle physics such as first observation of CP violation apart from the kaon, measurements of CKM matrix elements, measurements of purely leptonic B meson decays and searches for new physics.

In this work, we obtained mass spectrum and decay constants of the $D$ and $B$ mesons via QCD Sum Rule and a Quark Model potential. We also predicted decay constant for the $B_s$ meson where there is no specific experimental data exist. Harmonic oscillator wave function is used in the quark model and a sufficiently trivial interpolating current is used in QCD Sum Rule calculations. We studied ground states since they are accessible in the framework of QCD Sum Rules.

\section{\label{sec:level2}QCD Sum Rule Formalism}

In perturbation theory we assume that the eigenvalues and eigenfunctions can be expanded in a power series  as:
\begin{eqnarray*}
E_n=E_n^0+\lambda E_n^1 + \lambda^2 E_n^2+ \cdots \\
\vert n \rangle = \vert n^0 \rangle +  \lambda\vert n^1 \rangle + \lambda^2 \vert n^2 \rangle + \cdots ,
\end{eqnarray*}
where $n$ is the principal quantum number and $\lambda$ is a parameter. These series are in principle divergent, but they are asymptotic. This means that when the perturbation parameter is small, the first two or three terms are convergent so that the rest of the series can be ignored. In the case of QCD, due to the largeness of the parameter in lower energies, such a truncation, cannot be performed. The nonperturbative aspect of QCD makes almost impossible to study bound states in terms of perturbation theory. For this reason there is a need  of nonperturbative methods to overwhelm this situation and study bound states. Among others such as Effective Field Theory and Lattice QCD, QCD Sum Rule is maybe the most popular nonperturbative method. 

QCD Sum Rule is first formulated by Shifman, Vansthein and Zakharov for mesons in \cite{36} and generalized to baryons by Iofe in \cite{37}. The basic idea of the this formalism is to study bound state phenomena in QCD from the asymptotic freedom side, $i.e.$, to start evaluation of correlation function at short distances, where quark-gluon dynamics is perturbative and move to larger distances where hadronization occurs, including non-perturbative effects and using some approximate procedure to get information on hadronic properties \cite{38}. 

To obtain physical observables from QCD sum rules, a correlator of two hadronic currents which is defined as:

\begin{equation}
\Pi=i \int d^4 x e^{i p x} \langle 0 \vert {\cal T} j(x) j^\dagger(0) \vert 0 \rangle \label{eqn1},
\end{equation}
is studied. Here $p$ is momentum and $j(x) $ is a current composed of quarks and gluon fields with the hadron's quantum numbers. When this operator is applied to vacuum, it can create the hadron that we study. Eqn. (\ref{eqn1}) is known as correlation function. The fundamental assumption of the QCD sum rules is that there is a region of $p$ where correlation function can be equivalently described at both, quark and hadron sector. The former is known as QCD or OPE (Operator Product Expansion) side, and the latter is known as the phenomenological side. Matching these two sides of the sum rule, one can obtain information about hadron properties \cite{38}.

For $ p^2>0 $, resolution of identity operator of hadron states can be written between the operators. This results correlation function as:

\begin{equation}
\Pi=\sum_h \langle 0 \vert j \vert h(p) \rangle \frac{1}{p^2-m_h^2} \langle h(p) \vert j^\dagger \vert 0 \rangle \label{eq:piphen} + \mbox{higher states}. 
\end{equation}
It can be seen from Eqn. (\ref{eq:piphen}) shows the poles in the correlation function, which indicats the presence of hadrons, created by the operator  $j(x)$.

For $-p^2 \gg \Lambda_{QCD}^2 (p^2<0)$, major contribution to the correlation function will come from the $x \sim 0$ region \cite{39}. In this case the product of two operators can be written in terms of OPE:

\begin{equation}
{\cal T} j(x) j^\dagger(0) = \sum_d C_d(x) O_d \label{fr}.
\end{equation} 
Here $C_d(x)$  are the coefficients, which can be calculated by the perturbation theory, and $O_d's$ are the operators with the mass dimension $d$. If Fourier transformation applies to Eqn. (\ref{fr}), correlation function can be written as:
\begin{equation}
\Pi = \sum_d C^f_d(p) \frac{\langle O_d \rangle}{p^d} \label{eq:piqcd},
\end{equation}
where $\langle O_d \rangle$ are the vacuum condensates that cannot be calculated by perturbation theory except $d=0$. $d=0$ corresponds to unitary operator and can be calculated via perturbation theory. Other operators can be written as  $\langle \bar q  q\rangle$ ($d=3$); $m_q \langle \bar q q \rangle$ ($d=4$), $\langle G_{\mu \nu} G^{\mu \nu}\rangle$ ($d=4$), $\langle \bar q g \sigma G q\rangle$ ($d=5$). For  $d=1$ and $d=2$ there exist no operator. As a result of this, the expansion converges quickly although it is an infinite summation. 

In order to get sum rules we must equate Eqns. (\ref{eq:piphen}) and (\ref{eq:piqcd}). But these two expressions are obtained in different regions of $p$. By using spectral density representation of correlation function, this matching can be made:

\begin{equation}
\Pi(p^2) = \int_0^\infty \frac{\rho(s)}{s-p^2} + \mbox{ polynomials of $p^2$ } \label{eq:spektral}.
\end{equation}

Spectral density $\rho(s)$ can be acquired from Eqn. (\ref{eq:piphen}). Inserting $\rho(s)$ into Eqn.  (\ref{eq:spektral}), an expression of correlation function can be obtained from Eqn. (\ref{eq:piphen}) for $p^2<0$ region. If we denote $\rho^{phen}(s)$ as spectral density from Eqn. (\ref{eq:piphen}) and $\rho^{QCD}(s)$ from Eqn. (\ref{eq:piqcd}), we get
\begin{eqnarray}
\int_0^\infty ds \frac{\rho^{phen}(s)}{s-p^2} + \mbox{polynomials}&=& \int_0^\infty ds \frac{\rho^{QCD}(s)}{s-p^2} + \mbox{polynomials}.
\end{eqnarray}

In order to extract physical properties from this expression, one must eliminate the polynomial terms, for example by using derivatives. In principle no one knows the polynomial degree, and how many polynomials are. The correct procedure is then to use the Borel transformation, which contains infinite derivative:

\begin{equation}
{\cal B}_M^2\left[\Pi (q^2) \right]=\lim_{-q^2,n \to \infty, \quad \\ -q^2/n=M^2} \frac{-(q^2)^{n+1}}{n!}\left( \frac{d}{dq^2} \right)^n \Pi (q^2).
\end{equation}
 Here $M^2$ is defined as the Borel parameter \cite{36}. This transformation effectively removes the polynomials and makes 
 \begin{equation}
\frac{1}{s-p^2} \rightarrow e^{-\frac{s}{M^2}}.
\end{equation} 
Then:
\begin{equation}
\sum_h \vert \langle 0 \vert j \vert h(p) \rangle \vert^2 e^{-\frac{m_h^2}{M^2}} + \mbox{higher states} = \int_0^\infty \rho^{QCD}(s) e^{- \frac{s}{M^2}},
\label{eq:eq8}
\end{equation}
which resembles QCD parameters and hadronic properties. This equation still shows presence of unknown parameters. The $e^{-\frac{m_h^2}{M^2}}$ factor makes the contribution of small masses dominant. To parametrize contributions of higher states, {\ quark-hadron duality} approximation is used. According to quark-hadron duality, for $s>s_0$,  $\rho^{phen}(s) \simeq \rho^{QCD}(s)$.  $\rho^{phen}(s)$ has contribution of higher states and heavier hadrons when $s>s_0$. $s_0$ is called as continuum threshold, and is related mass of the hadron that is studied in sum rules. So, we can write Eqn. (\ref{eq:eq8}) as follows:
\begin{equation}
\vert \langle 0 \vert j \vert m_h(p) \rangle \vert^2 e^{-\frac{m_h^2}{M^2}} = \int_0^{s_0} \rho^{QCD}(s) e^{-\frac{s}{M^2}}.
\label{eq:eq9}
\end{equation}
In this equation  $m_h$ is the hadron of the smallest mass which can be created by $j$. 

Physical properties extracted from the sum rules must be independent of Borel parameter, ($M^2$). Here we assume that there exist a range of $M^2$, called Borel window, in which two sides have a good overlap and information on the lowest state can be extracted. Minimum and maximum values of Borel window can be extracted in a way that QCD side convergence gives the minimum value, and the condition that pole contribution should be bigger than the continuum contribution gives the maximum value of Borel window  \cite{38}. 

\subsection{Mass Sum Rule}

The mass sum rule can be obtained by matching QCD and phenomenological sides of correlation function \citep{36,38,39}. Here we will give the formula:
\begin{equation}
m^2=\frac{\int^{s_0}_{s_{min}} ds e^{-\frac{s}{M^2}} s \rho^{QCD}(s)}{\int^{s_0}_{s_{min}} ds e^{-\frac{s}{M^2}}  \rho^{QCD}(s)}.
\end{equation}

\subsection{Decay Constant}

The decay constant can be obtained from the formula \cite{40} as:
\begin{equation}
f^2_{m_h} = e^{\frac{m_h}{M^2}}\frac{1}{m_h^2} \int^{s_0}_{s_{min}} ds e^{-\frac{s}{M^2}}  \rho^{QCD}(s),
\end{equation}
where $m_h$ is the hadron mass extracted from sum rules. 

\section{Quark Model}
Also known as potential model or quark potential model, quark model considers one or more interacting particles under a given potential. In the early 60s quarks were modelled and experimental evidences were found subsequently. This approach provided a reliable basis to study and investigate particle physics and gave compatible results with the experiments.  

The most important part of the quark model is the potential. After the november revolution of particle physics in 1974, the year in which charmonium states were observed, new models were proposed to calculate spectrum and radiative transitions \citep{41,42,43}. The so-called Cornell potential, proposed in \cite{42}, reads as: 

\begin{equation}
V(r)=-\frac{\kappa}{r}+ar,
\end{equation}

where $\kappa$ and $a$ are some parameters to extract from fit to the experimental data . This potential is still used with some modifications to account for example hyperfine splittings in the energy levels. The other potentials such as power law potential \cite{44}, logarithmic potential \cite{45}, Richardson potential \cite{46}, Buchmüller-Tye potential \cite{47}, and Song-Lin potential \cite{48} were used to fit quarkonium spectra, and gave good results in agreement with experiments. These were phenomenological spin-independent potentials and not directly QCD motivated. The interquark potential was not derived from first principles of QCD in the early quarkonium phenomenology. This means, in terms of QCD, that potential is universal  (flavour independent) and since quarks are colorless particles, it was reasonable to assume the universality as valid, despite the fact that gluons couple to color charge. These spin-independent potential models performed good but not complete explanation of the energy level splittings. If we want to accommodate these splittings in the theory, we have to take care $i.e.$ of {\it spin-spin} and {\it spin-orbit} interactions in the model. \cite{49} reports an example of a QCD-motivated, spin- and velocity-dependent potential. These potentials deliver reliable results.

\section{Elaboration of the Problem}

\subsection{QCD Sum Rules}
In QCD sum rules,  the choice of the $j(x)$ current is important, since it creates hadrons from vacuum. We used the current:

\begin{equation}
j(x)=i\bar{Q}_a(x)\gamma_5 q_a(x),
\end{equation} 

where $Q$ is heavy quark, $q$ is light quark, $a$ is the color index, and $\gamma_5$ is the Dirac matrix. We take care of $m_q \to 0$ limit. In the limit of $m_q \to 0$, there appears a flavor symmetry between $b$ and $c$ quarks. By this symmetry it is possible to extract information about $c$ and $b$ sector with the same current. $b$ and $c$ quarks are heavy quarks so that it cannot be expected to be in the vacuum by themselves. So it is possible to ignore such condensate terms like $\langle \alpha_s \frac{GG}{\pi} \rangle$ and $\langle \bar{q} g_s \sigma G q \rangle $.  By introducing the current term into the Eqn. (\ref{eqn1}), one can obtain the following spectral density:

\begin{eqnarray}
F(s_0,M^2)&=&-\langle q\bar{q} \rangle e^{-\frac{m_Q^2}{M^2}}m_Q \nonumber \\
 &+& 6 e^{-\frac{s_0}{M^2}} e^{-\frac{s_u}{{M^2}}} (u_1-u_2) \nonumber \\ & \times & 
 [  e^{-\frac{s_0}{{M^2}}} M^2 (m_Q^2+s(u)+M^2) \nonumber \\ 
 &-& e^{-\frac{s_u}{{M^2}}} (m_Q^4+M^2(s_0+M^2))  ] \label{fsm}
\end{eqnarray}

where $\langle q\bar{q} \rangle$ is the condensate, and $u_1$ and $u_2$ are solutions of $s(u)=\frac{m_Q^2}{1-u}+\frac{m_q^2}{u}=s_0$. 

The mass sum rule can be obtained by taking derivative with respect to $1/M^2$, and dividing the result by Eqn. (\ref{fsm}): 

\begin{equation}
m_h^2=M^4 \frac{1}{F(s_0,M^2)} \frac{dF(s_0,M^2)}{dM^2}.
\end{equation}
The decay constant sum rule can be obtained as:
 
 \begin{equation}
f^2_{m_h}=e^{\frac{m_h^2}{M^2}} \frac{1}{m_h^2} F(s_0,M^2).
 \end{equation}

\begin{figure}[H]
\centering
\includegraphics[width=3.4in]{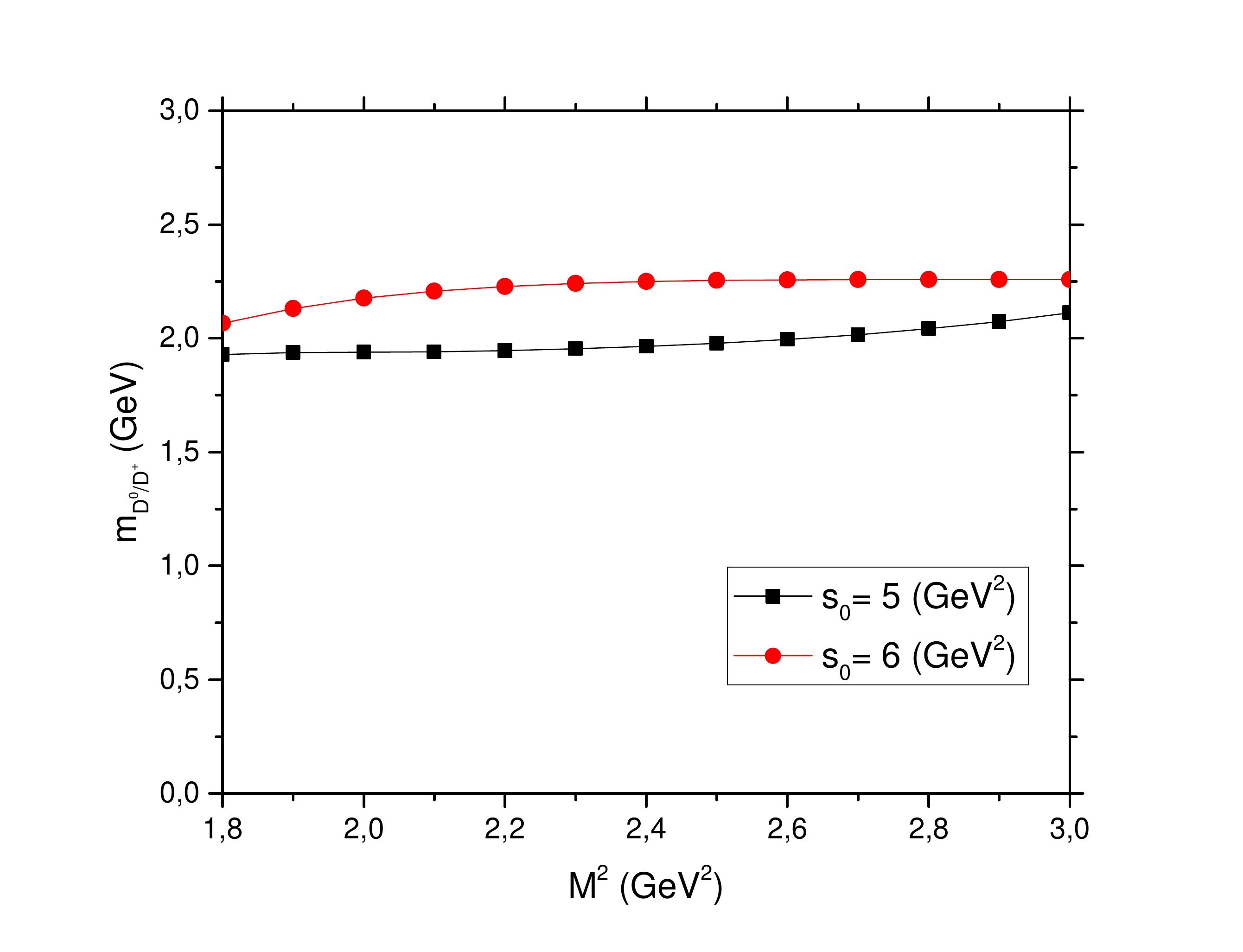}% Here is how to import EPS art
\caption{\label{fig:1} Borel parameter dependence of the  $D^0/D^+$ masses}
\end{figure}

\begin{figure}[H]
\centering
\includegraphics[width=3.4in]{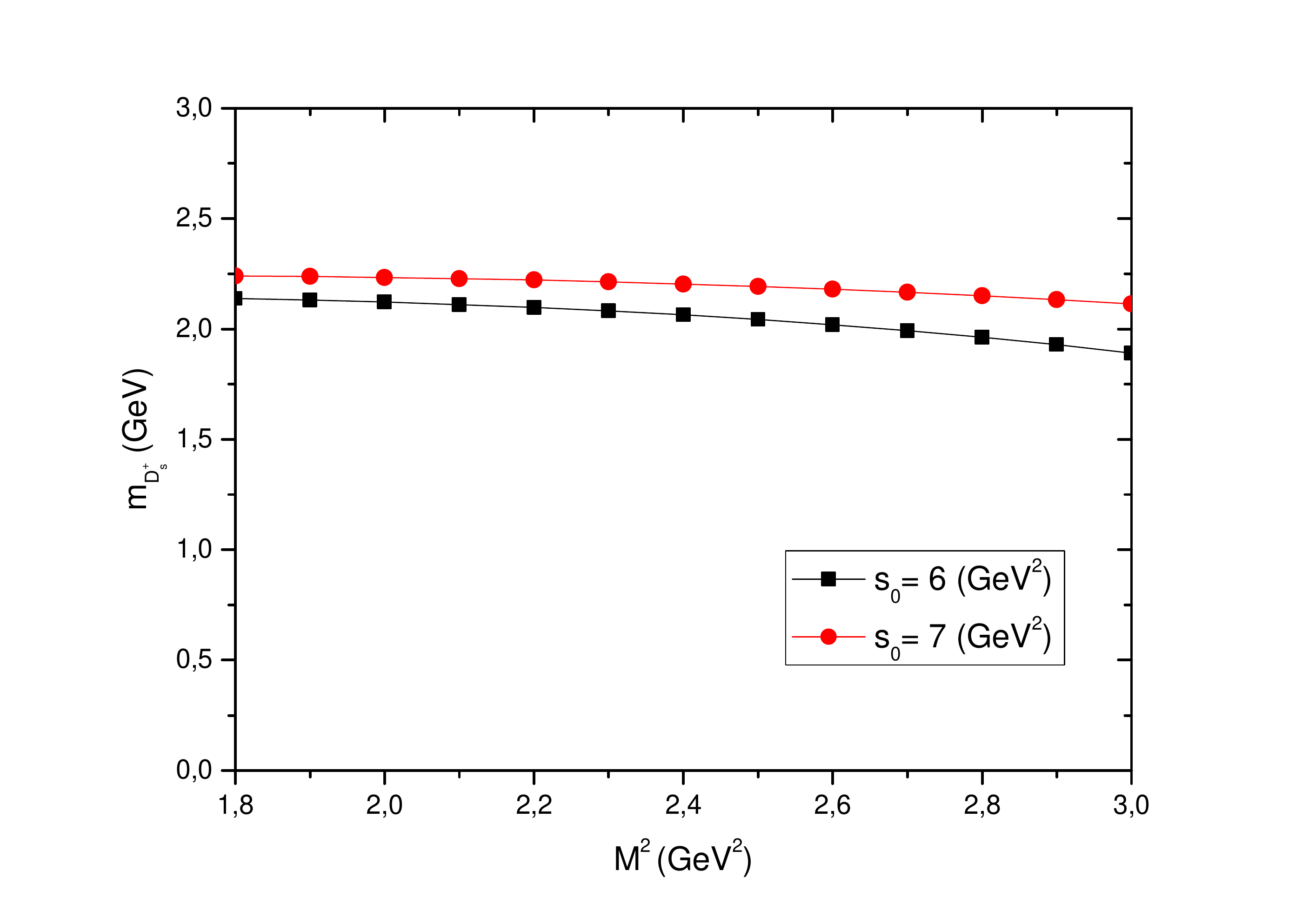}% Here is how to import EPS art
\caption{\label{fig:2} Borel parameter dependence  of the $D_s^+$ mass}
\end{figure}

\begin{figure}[H]
\centering
\includegraphics[width=3.4in]{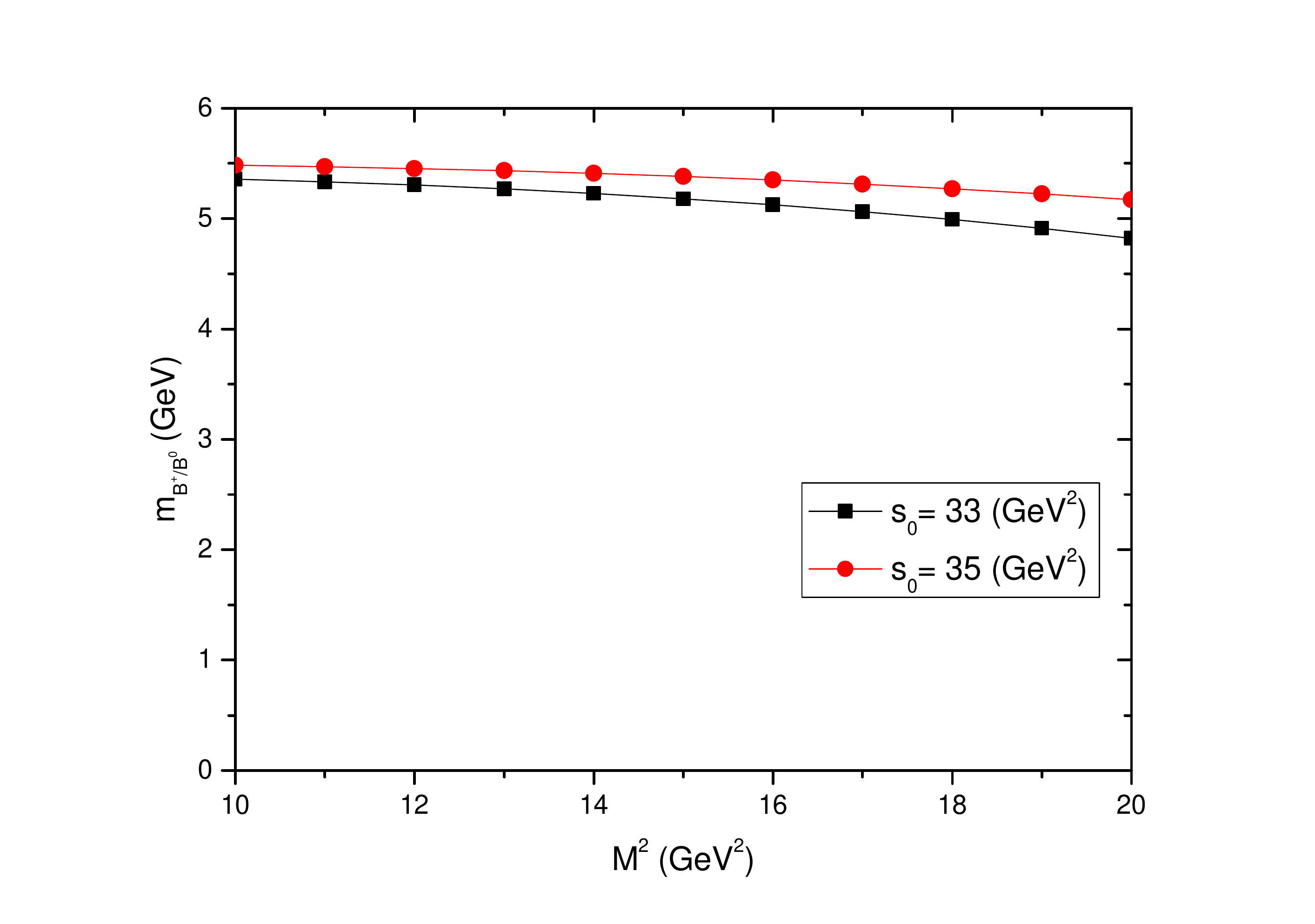}% Here is how to import EPS art
\caption{\label{fig:3} Borel parameter dependence of the $B^+/B^0$ masses}
\end{figure}

\begin{figure}[H]
\centering
\includegraphics[width=3.4in]{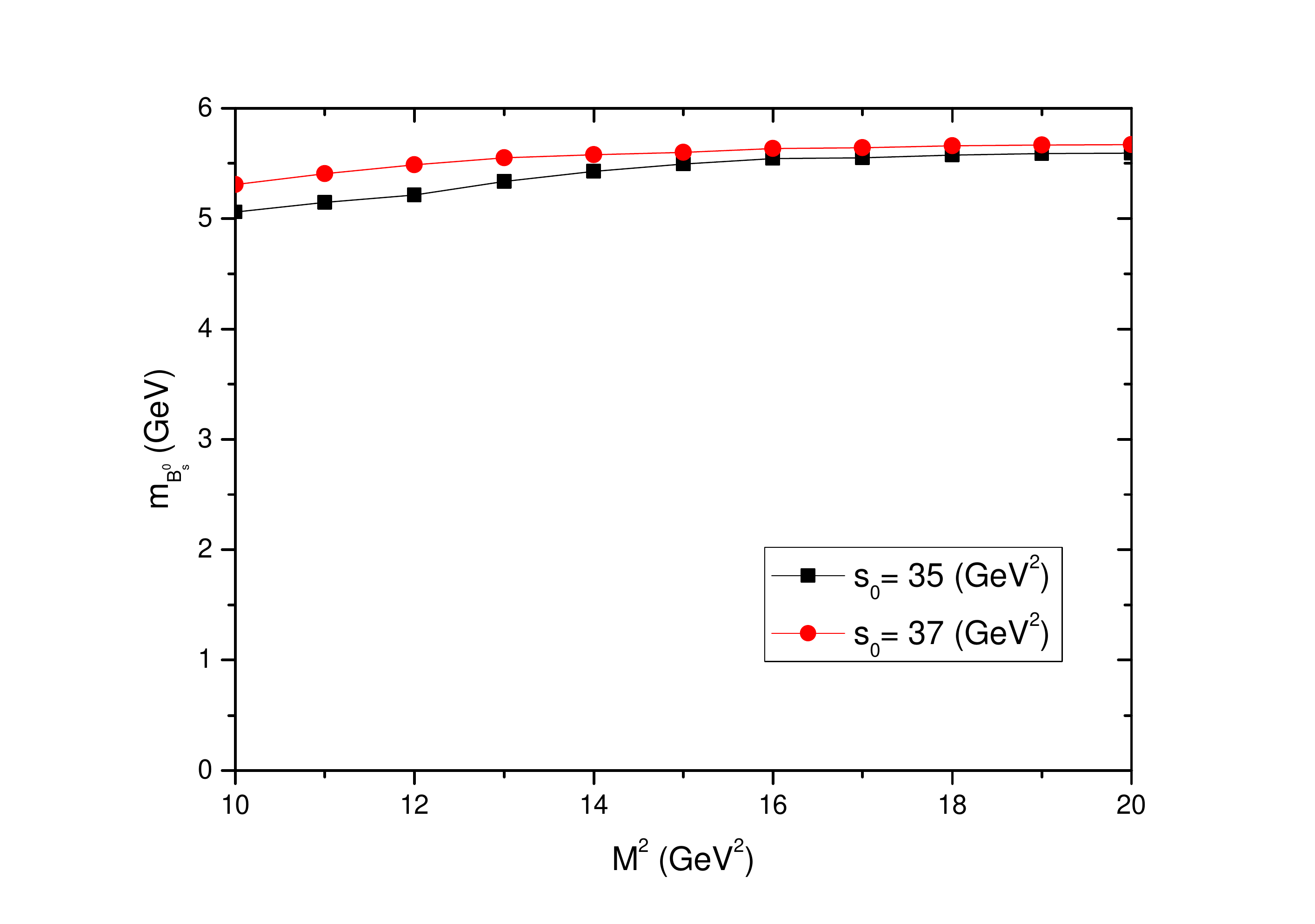}% Here is how to import EPS art
\caption{\label{fig:4} Borel parameter dependence of the $B_s^0  $ mass}
\end{figure}

\begin{figure}[H]
\centering
\includegraphics[width=3.4in]{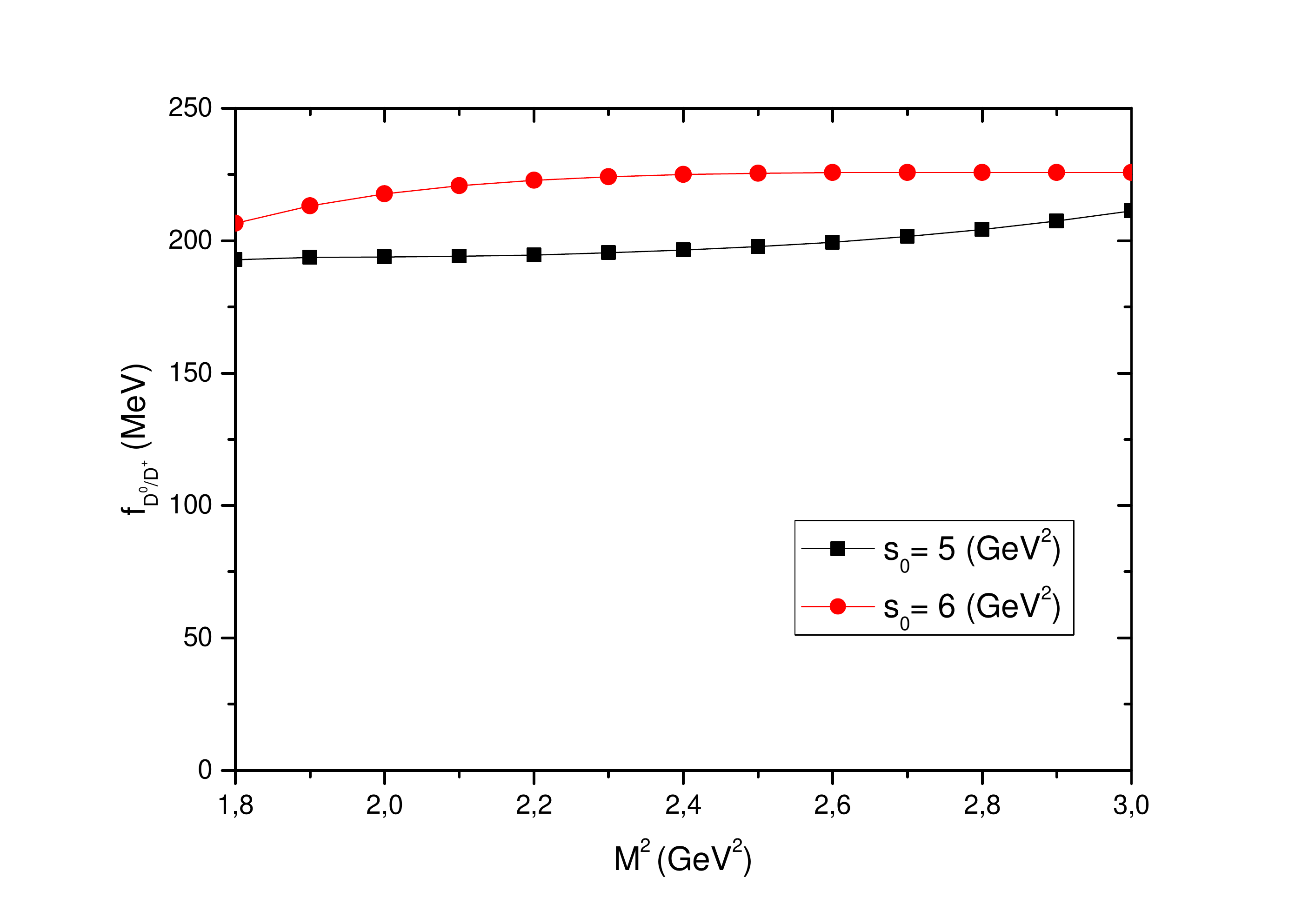}% Here is how to import EPS art
\caption{\label{fig:5} Borel parameter dependence of the $D^0/D^+$ decay constants}
\end{figure}

\begin{figure}[H]
\centering
\includegraphics[width=3.4in]{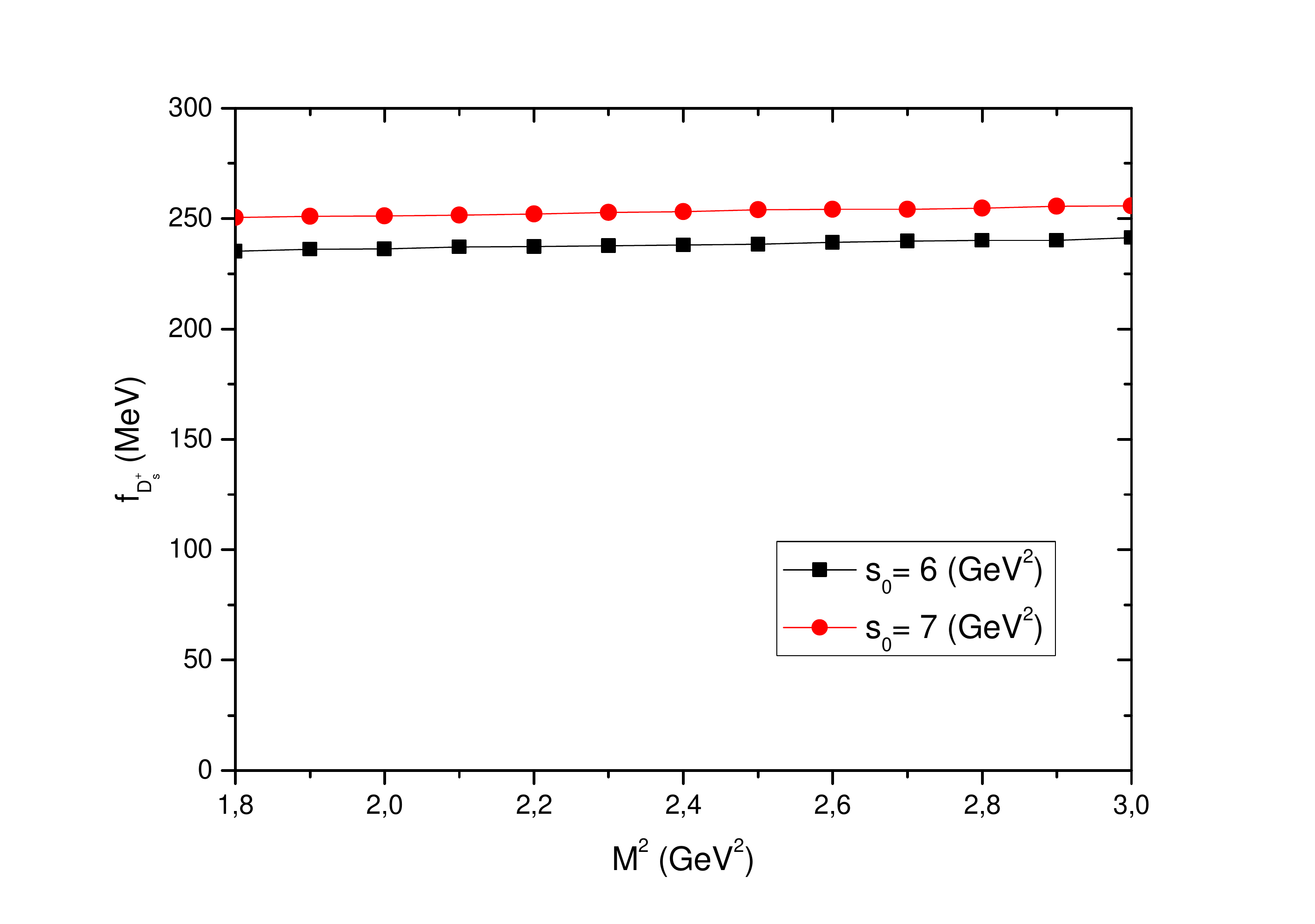}% Here is how to import EPS art
\caption{\label{fig:6} Borel parameter dependence of the $D_s^+$ decay constant}
\end{figure}

\begin{figure}[H]
\centering
\includegraphics[width=3.4in]{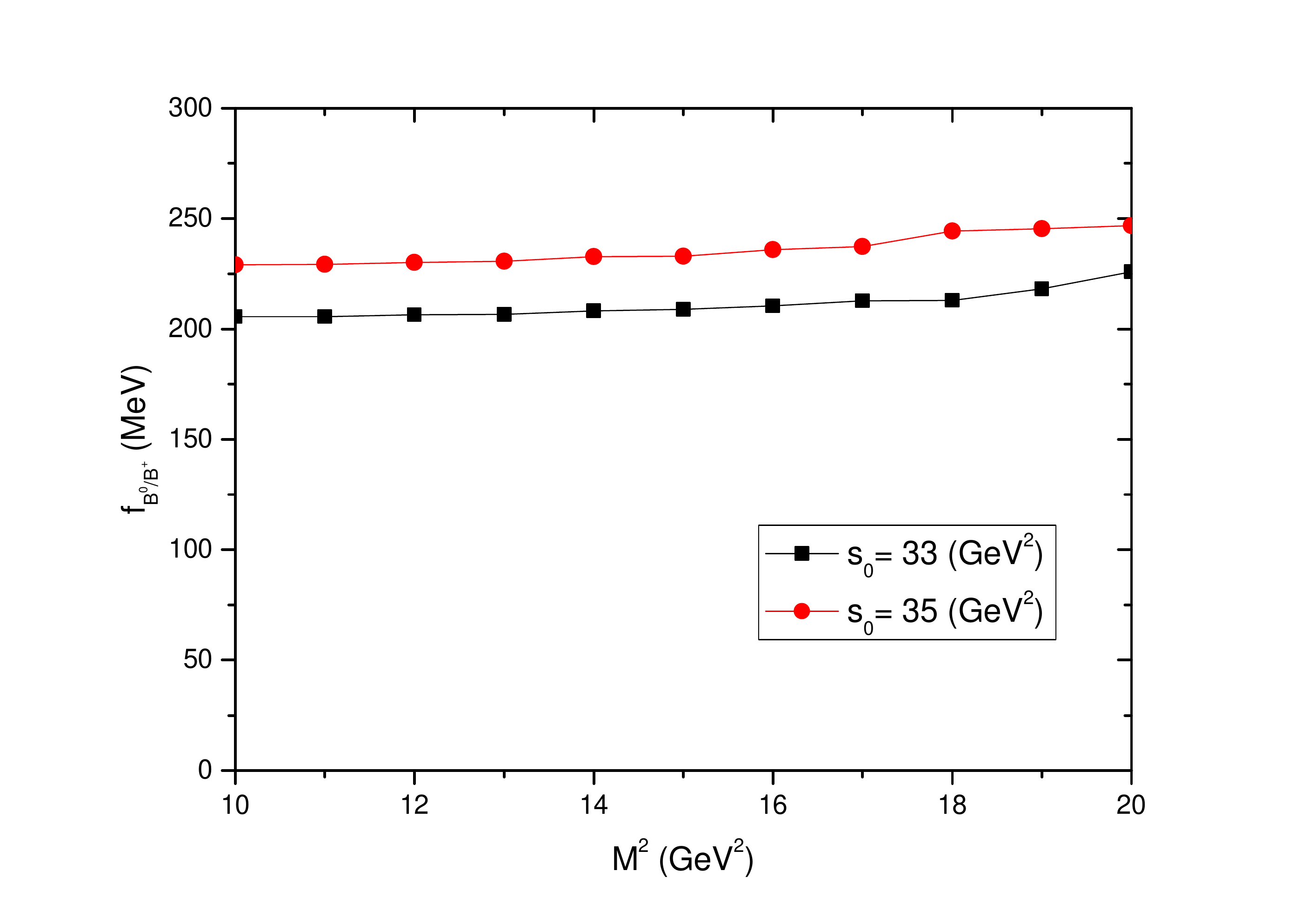}% Here is how to import EPS art
\caption{\label{fig:7} Borel parameter dependence  of the $B^+/B^0$ decay constants}
\end{figure}

\begin{figure}[H]
\centering
\includegraphics[width=3.4in]{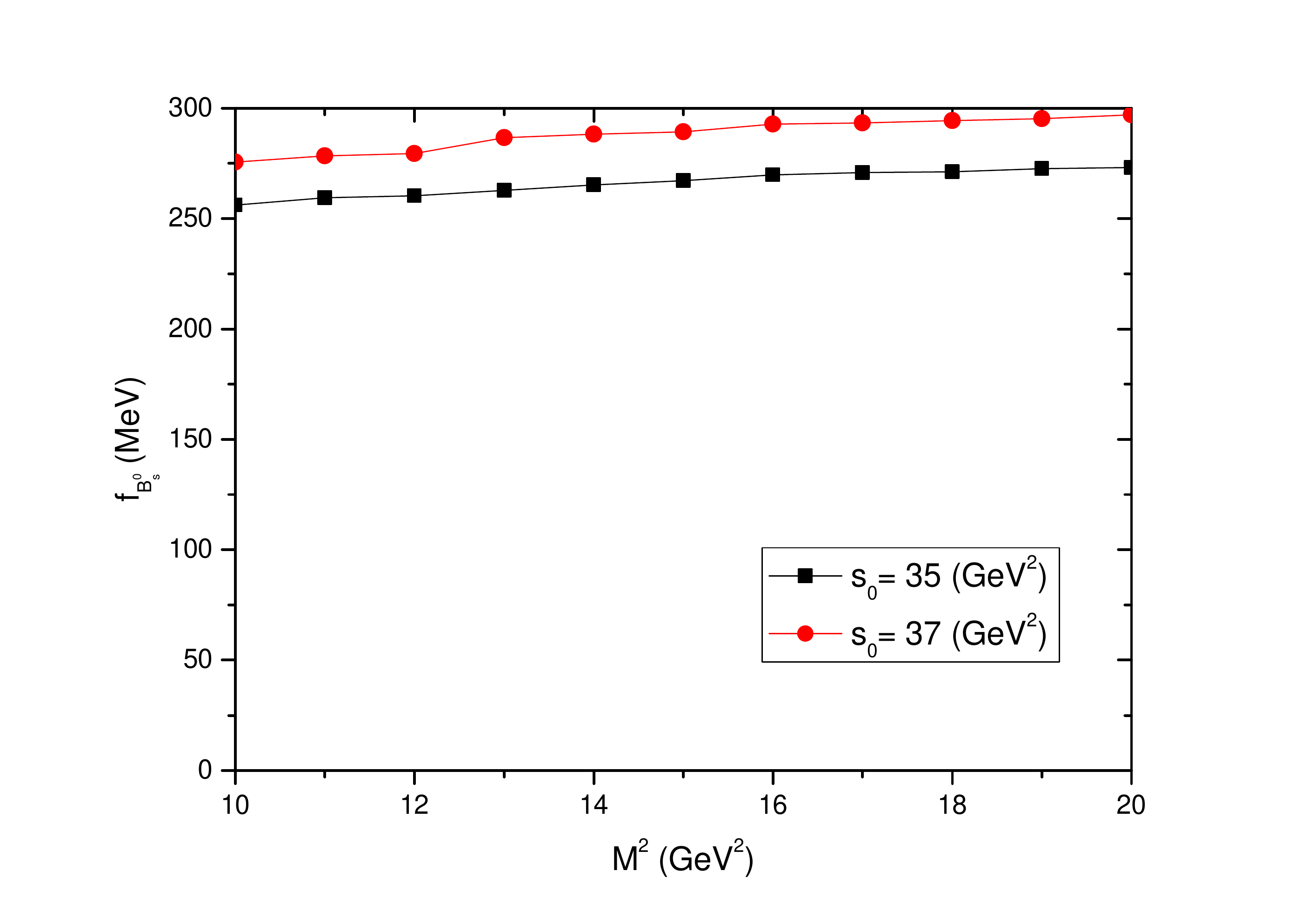}% Here is how to import EPS art
\caption{\label{fig:8} Borel parameter dependence of the $B_s^0  $ decay constant}
\end{figure}

The mass values and decays constants for heavy-light mesons are presented in Table \ref{tab:table1} and \ref{tab:table2}  and  Figs. \ref{fig:1}-\ref{fig:8}. 

\subsection{Quark Model}

Energy eigenvalues can be obtained by solving the Schrödinger equation in the quark model. The Schrödinger equation reads as:

\begin{equation}
H\vert \Psi_n \rangle =E_n \vert \Psi_n \rangle \label{sch},
\end{equation}

where $n$ denotes the principal quantum number. We can separate the wave function into radial $R_{nl}$ and angular parts $Y_{lm}(\theta, \phi)$ as follows:

\begin{equation}
\Psi_{nlm}(r, \theta, \phi )=R_{nl}(r) Y_{lm}(\theta, \phi).
\end{equation}

$R_{nl}$ is the radial wave function given as:

\begin{equation}
R_{nl}=N_{nl} r^l e^{-\nu r^2} L^{l+\frac{1}{2}}_{ \frac{n-l}{2}} \left( 2\nu r^2 \right) \label{eq:psir},
\end{equation}
with the associated Laguerre polynomials $L^{l+\frac{1}{2}}_{ \frac{n-l}{2}}$ and the normalization constant:

\begin{equation}
N_{nl}=\sqrt{\sqrt{ \frac{2\nu^3}{\pi}} \frac{2 (\frac{n-l}{2})!\nu^l}{(\frac{n+l}{2}+1)!!}}.
\end{equation}

With the wave function in hand one can obtain masses as well as decay constants for heavy and light mesons. The mass spectra can be obtained by solving Eqn. (\ref{sch}). For the decay constants we employ the following formulas,

which result are:
\begin{eqnarray}
f_p &=&\sqrt{\frac{3}{m_p}}  \times \int \frac{d^3k}{(2\pi)^3} \sqrt{1+\frac{m_q}{E_k}} \times \sqrt{1+\frac{m_{\bar{q}}}{E_{\bar{k}}}}  \nonumber \\ &\times& 
 \left( 1-\frac{k^2}{(E_k+m_q)(E_{\bar{k}}+m_{\bar{q}})} \right) \phi (\vec{k}),
\end{eqnarray}
for pseudoscalar mesons; and

\begin{eqnarray}
f_v &=&\sqrt{\frac{3}{m_v}} \times \int \frac{d^3k}{(2\pi)^3}  \sqrt{1+\frac{m_q}{E_k}} \times \sqrt{1+\frac{m_{\bar{q}}}{E_{\bar{k}}}}  \nonumber \\ &\times& 
  \left( 1+\frac{k^2}{3(E_k+m_q)(E_{\bar{k}}+m_{\bar{q}})} \right) \phi (\vec{k}),
\end{eqnarray}
for the vector mesons \cite{50}.  

In the nonrelativistic limit, these two equations take a simple form, which is known to be Van Royen and Weisskopf relation \cite{51}. For the meson decay constants:

\begin{equation}
f^2_{p/v}=\frac{12 \vert \Psi_{p/v}(0) \vert^2}{m_{p/v}}.
\end{equation}
Here $m_{p/v}$ denotes the pseudoscalar and vector mass of the related meson.

The results are shown in Tables \ref{tab:table1} and \ref{tab:table2}.

\begin{table}[H]
\caption{\label{tab:table1}Mass spectra of heavy-light mesons in MeV. QM denotes quark model and SR denotes sum rule calculations. The parameters are $\kappa=0.471$, $a=0.192 ~ GeV^2$, $m_c=1.320 ~ GeV$, $m_b=4.740 ~ GeV$ \cite{52}, $\langle \bar{q}q \rangle=0.241  ~ GeV^3$, $m_u=m_d=0.340 ~ GeV$ and $m_s=0.600 ~ GeV$.}
\begin{ruledtabular}
\begin{tabular}{ccccccc}
Meson & Exp. \cite{2}& QM & SR & \cite{9} & \cite{14} & \\
\hline
$D^0/D^+$& 1869.3 $\pm$ 0.4  & 1859 & 1972 $\pm$ 94 & 1870.82 & 1854.7 \\
$D_s^+$  & 1968.2 $\pm$ 0.4  & 2056 & 2118 $\pm$ 75 & 1966.62 & 1974.5 \\
$B^+/B^0$& 5279.0 $\pm$ 0.5  & 5260 & 5259 $\pm$ 109& 5273.50 & 5277.2 \\
$B_s^0  $& 5367.7 $\pm$ 1.8  & 5442 & 5488 $\pm$ 76 & 5365.99 & 5384.8 \\
\end{tabular}
\end{ruledtabular}
\end{table}

\begin{table}[H]
\caption{\label{tab:table2}Pseudoscalar and vector decay constants of heavy-light mesons in MeV. QM denotes quark model and SR denotes sum rule calculations. }
\begin{ruledtabular}
\begin{tabular}{cccccccccc}
Meson & Exp. & QM & SR & \cite{9} & \cite{16} & \cite{53} \\
\hline
$D^0/D^+$& 206 $\pm$ 8.9 & 199 & 210.25 $\pm$ 11.60 & 205.14 & 206.2 $\pm$ 7.3 $\pm$ 5.1 &  207.53 \\
$D_s^+$  & 249 & 253 & 245.70 $\pm$ 7.46  & 241.84 & 245.3 $\pm$ 15.7 $\pm$ 4.5 & 262.56 \\
$B^+/B^0$& 204 $\pm$ 31 & 209 & 223.45 $\pm$ 12.4  & 201.09 & 193.4 $\pm$ 12.3 $\pm$ 4.3 & 208.13 \\
$B_s^0  $&  & 275 & 277.22 $\pm$ 11    & 292.04 & 232.5 $\pm$ 18.6 $\pm$ 2.4 & 262.39 \\
\end{tabular}
\end{ruledtabular}
\end{table}

\section{Summary and Conclusions }

In this paper, we calculated mass spectra and decay constants of pseudoscalar and vector heavy-light mesons ($B$, $B_s$, $D$ and $D_S$) in the framework of QCD sum rule and quark model. Obtained results for masses of $B$ and $D$ mesons are in good agreement with the available experimental data. In the mass spectra, the extrapolation via quark model gave close results to experimental data than the  QCD sum rule consideration. The QCD sum rules approach gives reasonable but not very good-matching results compared to the experimental values, because of the adopted approximation when evaluating the current, whereas the higher dimension of that operator could improve the estimates. Other potentials and further studies should be taken in consideration for a better understanding. 

The heavy-light mesons under study in this paper are well established indeed, and any prediction or reproduction of mass spectrum does not directly guarantee the validity of the model, but shows a possible path to follow for a further investigation. Therefore other physical observables such as decay constants should be experimentally investigated to give more inputs to the theory. For example the only precise value of decay constant is known for $D$ mesons, and systematics are evaluated. The other mesons in this study need more experimental data.  For $B_s$ there is no available experimental data. We predicted for the first time decay constant value for $B_s$ in this manner. 

Decay constants give information about short distance structure of hadrons. The obtained results for decay constants are in agreement with the other studies and available data. We did not consider in this work relativistic corrections. 

In QCD Sum Rule calculations, physical observables must be independent of the Borel parameter. In Figs. \ref{fig:1}-\ref{fig:8} the smoothness of the graphs are compatible with existing data. It is worthy to note that in Fig. \ref{fig:1} and Fig. \ref{fig:5} the 'slope' of the two curves of $D^0/D^+$ are not in the same range. The reason for that could be the smallness of the Borel parameter and continuum threshold energy, since correlation function receives main contribution at $s \neq M^2$. On the other hand, the smallness of Borel parameter can blow up the corrections to the perturbative part of the correlation function. 

In summary, we obtained good results in accordance with the available data and theoretical studies. As mentioned before, other potential models and interpolating currents can be used to study mass spectra and decay constants. Heavy-light systems in view of the quark model are important to study hadronic interactions. Especially Heavy Quark Spin Symmetry can play an essential role in heavy-light systems. The higher dimensions of the operators in interpolating currents would deliver more accurate results.


\begin{thebibliography}{}
\bibitem{1}
M. Gell-Mann, A schematic model of baryons and mesons, Phys. Lett.\textbf{8}, 214  (1964).

\bibitem{2}
C. Patrignani et al. (Particle Data Group), Chin. Phys. C \textbf{40} 100001 (2016) and 2017 update.

\bibitem{3}
Simon Capstick, Stephen Godfrey, Nathan Isgur, and Jack E. Paton, Taking the ’naive’ and
’nonrelativistic’ out of the quark potential model, Phys. Lett., B \textbf{175} 457, (1986).

\bibitem{4} 
 S. Gasiorowicz and J. Rosner, Hadron spectra and quarks, Am. J. Phys \textbf{49}, 954, (1981).

\bibitem{5}
 B. Julia- Diaz and D. Riska, Baryon magnetic moments in relativistic quark models, Nucl. Phys. A \textbf{739}, 69-88 (2004).
 
\bibitem{6}
K. B. V. Kumar, B. Hanumaiah, S. Pepin, Meson spectrum in a relativistic harmonic model with instanton-induced interaction, Eur. Phys. J. A \textbf{19}(2), 247-250 (2004).

\bibitem{7}
J-M. Richard, An Introduction to Quark Model, arXiv:[hep-ph] 1205.432v2

\bibitem{8}
D. Ebert, R. N. Faustov, V. O. Galkin, Radiative M1-decays of heavy-light mesons in the relativistic quark model, Phys. Lett. B \textbf{537} 241-248 (2002).

\bibitem{9}
K. K. Pathak, D. K. Choudhury, N. S. Bordoloi, Leptonic decay of Heavy-light Mesons in a QCD Potential, Int. J. Mod. Phys. A \textbf{28}(2), 1350010 (2013).

\bibitem{10}
D. Ebert,  R. N. Faustov, V. O. Galkin, Decay Constants of Heavy-Light Mesons in the Relativistic Quark Model, Mod. Phys. Lett. A \textbf{17}(3) 803-807 (2002).

\bibitem{11}
T. A. Lahde, C. J. Lyfalt, D. O. Riska, Spectra and M1 Decay Widths of Heavy-Light Mesons, Nucl. Phys. A \textbf{674}(1-2) 141-167 (2000).

\bibitem{12}
D. U. Matrasulov, F. C. Khanna, H. Yusupov, Spectra of Heavy-Light Mesons, J. Phys. G: Nucl. Part. Phys. \textbf{29} 475-483 (2003)

\bibitem{13}
B. L. G. Bakker, Spectrum and decay donstants of heavy-light mesons, Few Body Sys. \textbf{44} 91-93 (2008).

\bibitem{14}
B. H. Yazarloo, H. Mehraban, Mass spectrum and decay properties of heavy-light mesons: $D$, $D_s$, $B$ and $B_s$ mesons, Eur. Phys. J. Plus \textbf{132} 80 (2017).

\bibitem{15}
J.-B. Liu, M-Z. Yang, Heavy-light mesons in a relativistic model, Chin. Phys. C \textbf{40}7, 073101 (2016).

\bibitem{16}
W. Lucha, D. Melikhov, S. Simula, Decay constants of heavy pseudoscalar mesons from QCD sum rules, J. Phys. G: Nucl. Part. Phys. \textbf{38}10, 105002 (2011).

\bibitem{17}
J.-B. Liu, C.-D. Lü, Spectra of heavy–light mesons in a relativistic model, Eur. Phys. J. C  \textbf{77} 312 (2017).

\bibitem{18}
T. Lesiak, B Meson Spectroscopy, Acta Phys. Pol. B \textbf{29} 3379-3386 (1998).

\bibitem{19}
H. A. Alhendi, T. M. Aliev, M. Savci, Strong decay constants of heavy tensor mesons in light cone QCD sum rules, J. High Energy Phys. \textbf{4} 050 (2016).

\bibitem{20}
Z. G. Wang, Analysis of the masses and decay constants of the heavy-light mesons with QCD sum rules,  Eur. Phys. J. C  \textbf{75} 427 (2015).

\bibitem{21}
A. K. Rai, R. H. Parmar, P. C. Vinodkumar, Masses and decay constants of heavy-light flavour mesons in a variational scheme, J. Phys. G: Nucl. Part. Phys. \textbf{28}8, 2275-2282  (2002).

\bibitem{22}
T. Huang, Decay constants of heavy-light mesons in heavy quark effective theory, Phys. Rev. D \textbf{53}9  5042-5050 (19969.

\bibitem{23}
A. Duncan, E. Eichten, J. M. Flynn,  B. R. Hill, H. Thacker, Masses and Decay Constants of Heavy-Light Mesons Using the Multistate Smearing Technique, Nucl. Phys. B \textbf{34} 444-452 (1994).

\bibitem{24}
P. Gelhausen, A.  Khodjamirian, A. A. Pivovarov,  D. Rosenthal, Decay constants of heavy-light vector mesons from QCD sum rules, Phys. Rev. D \textbf{88} 014015 (2013). Erratum (\textbf{91} 099901 (2015)) Erratum (\textbf{89} 099901 (2014)).

\bibitem{25}
S. Narison, Decay Constants of Heavy-Light Mesons from QCD, Nucl. Part Phys. P \textbf{270} 143-153 (2016). 

\bibitem{26} 
H.-y. Cheng, X.-W. Kang, Branching fractions of semileptonic $D$ and $D_s$ decays from the covariant light-front quark model, Eur. Phys. J. C \textbf{77}(9) 587 (2017).

\bibitem{27}
 X.-W. Kang, B. Kubis, C. Hanhart, U.-G., Meissner, $B_{l4}$ decays and the extraction of $\vert V_{ub} \vert$, Phys. Rev. D \textbf{89} 053015 (2014).


\bibitem{28}
N. Brambilla, S. Eidelman, P. Foka et al.,  QCD and strongly coupled gauge theories: challenges and perspectives, Eur. Phys. J. C  \textbf{74}(10) 2981 (2014).

\bibitem{29}
B. Aubert et al. (BABAR Collaboration), Observation of a New $D_s$ Meson Decaying to $DK$ at a Mass of $2.86 ~ {\rm GeV}/c^2$ , Phys. Rev. Lett. \textbf{97}, 222001 (2006).

\bibitem{30}
J. Brodzicka et al. (Belle Collaboration), Observation of a New $D_{sJ}$ Meson in $B^+ \to \bar{D}^0 D^0 K^+$ Decays, Phys. Rev. Lett. \textbf{100}, 092001 (2008).

\bibitem{31}
R. Aaij et al. (LHCb Collaboration), Dalitz plot analysis of $B_s^0 \to \bar{D}^0 K^- \pi^+ $ decays, Phys. Rev. D \textbf{90}, 072003 (2014).

\bibitem{32}
R. Aaij et al. (LHCb Collaboration), Observation of Overlapping Spin-1 and Spin-3 $ \bar{D}^0 K^-$ Resonances at Mass $ 2.86 ~ {\rm GeV}/c^2$, Phys. Rev. Lett. \textbf{113}, 162001 (2014).

\bibitem{33}
N. Cabibbo, Unitary Symmetry and Leptonic Decays, Phys. Rev. Lett. \textbf{10}, 531-533  (1963).

\bibitem{34}
J. H. Christenson, J. W. Cronin, V. L. Fitch, and R. Turlay, Evidence for the 2$\pi $ Decay of
the K02 Meson, Phys. Rev. Lett. \textbf{13}, 138-140 (1964).

\bibitem{35}
M. Kobayashi and T. Maskawa, CP Violation in the Renormalizable Theory of Weak Interaction,  Prog. Theor. Phys. \textbf{49}, 652-657 (1973) .

\bibitem{36}
M. A. Shifman, A. I. Vainsthein and  V. I. Zakharov, QCD and Resonance Physics, Nucl. Phys. B \textbf{147} 447-518 (1979).

\bibitem{37}
 B. L. Ioffe,  Calculation of Baryon Masses in Quantum Chromodynamics, Nucl. Phys. B \textbf{188}(2)  317-341 (1981).
 
\bibitem{38}
M. Nielsen, F. S. Navarra, S. H. Lee, New charmonium states in QCD sum rules: A concise review, Phys. Rep. \textbf{497} 41-83 (2010). 

\bibitem{39}
P. Colangelo,  A. Khodjamirian, QCD sum rules, a modern perspective, In  M. Shifman (Ed.): At the frontier of particle physics, vol. 3 1495-1576 (arXiv:hep-ph/0010175)

\bibitem{40}
H. Sundu, The Mass and Current-Meson Coupling Constant of the Exotic $X(3872)$ State from QCD Sum
Rules, SDU J. Nat. Appl. Sci. \textbf{20}(3) 448 (2016).

\bibitem{41}
A. De Rújula and S. L. Glashow, Is bound charm found? Phys. Rev. Lett. \textbf{34} 46–49 
(1975).

\bibitem{42}
E. Eichten, K. Gottfried, T. Kinoshita, J. Kogut, K. D. Lane, and T. M. Yan,  Spectrum of
charmed quark-antiquark bound states, Phys. Rev. Lett. \textbf{34} 369–372 (1975).

\bibitem{43}
Thomas Appelquist, A. De Rújula, H. David Politzer, and S. L. Glashow,  Spectroscopy of
the new mesons,  Phys. Rev. Lett. \textbf{34} 365–369  (1975).

\bibitem{44}
A. Martin, A Fit of Upsilon and Charmonium Spectra, Phys. Lett. B. \textbf{93}, 338 (1980).

\bibitem{45}
C. Quigg, J. L. Rosner, Quarkonium level spacings, Phys. Lett. B \textbf{71}(1) 153-157 (1977).

\bibitem{46}
J. L. Richardson, The heavy quark potential and the $\Upsilon, J/\psi$ systems,  Phys. Lett. B \textbf{82}(2) 272-274 (1979).

\bibitem{47}
W. Buchmüller and S. -H. H. Tye, Quarkonia and quantum chromodynamics, Phys. Rev. D \textbf{24} 132 (1981).

\bibitem{48}
Song Xiaotong and Lin Hefen, A New Phenomenological Potential for Heavy Quarkonium, Z. Phys. C - Particles and Fields, \textbf{34}, 223-231 (1987). 

\bibitem{49}
E. Eichten and F. Feinberg, Spin-dependent forces in quantum chromodynamics, Phys. Rev. D \textbf{23} 2724 (1981).

\bibitem{50}
O. Lakhina and E. S. Swanson, Dynamics properties of charmonium,  Phys. Rev. D \textbf{74} 014012 (2006).

\bibitem{51}
R. Van Royen and V. F. Weisskopf, Nuovo Cimento A \textbf{50}, 617 (1967)

\bibitem{52}
S. Jacobs, M. G. Olsson and Casimir Suchyta III, Comparing the Schrödinger and spinless Salpeter equations for heaavy quark bound states, Phys. Rev. D \textbf{33}(11) 3338-3348 (1996).


\bibitem{53}
Seung-il Nam, Extended nonlocal chiral-quark model for the $D-$ and $B-$ meson weak-decay constants, Phys. Rev. D \textbf{85} 034019 (2012).






\end{thebibliography}
\end{document}